\documentclass[twocolumn,10pt]{article}
\usepackage[top=2cm, bottom=2cm, left=1.75cm, right=1.75cm]{geometry}
\usepackage{times}

\usepackage{bbold}
\usepackage{amsmath}
\usepackage{flushend}
\usepackage{algorithm}
\usepackage{algpseudocode}
\usepackage[dvipsnames]{xcolor}
\usepackage[hyphens]{url}
\usepackage{paralist}
\usepackage{graphicx}		
	\setkeys{Gin}{width=\textwidth, height=\textheight, keepaspectratio}

\usepackage{booktabs}
\usepackage[square,sort&compress,comma,numbers]{natbib}
\usepackage[labelfont=bf]{caption}
\usepackage{amssymb}
\usepackage{pifont}
\usepackage{multirow}
\usepackage{tablefootnote}

\usepackage[bookmarks=true, bookmarksnumbered=true, colorlinks=true, linkcolor=linkcol, citecolor=citecol, urlcolor=urlcol, hypertexnames=true]{hyperref}

\definecolor{darkgreen}{RGB}{0, 100, 0}
\definecolor{linkcol}{rgb}{0.3,0,0}
\definecolor{citecol}{rgb}{0.3,0,0}
\definecolor{urlcol}{rgb}{0.3,0,0}
\definecolor{vlightgray}{gray}{0.925}

\let\OLDthebibliography\thebibliography
\renewcommand\thebibliography[1]{
  \OLDthebibliography{#1}
  \setlength{\parskip}{0pt}
  \setlength{\itemsep}{1pt plus 0.2ex}
}

\usepackage[hang,flushmargin]{footmisc}

\makeatletter
\def\url@leostyle{%
  \@ifundefined{selectfont}{\def\UrlFont{}}%
  {\def\UrlFont{}}%
}
\makeatother
\usepackage[hyphenbreaks]{breakurl}

\usepackage[raggedright,compact]{titlesec}
\titlespacing*{\section}{0pt}{*3.5}{3.5pt}
\titlespacing*{\subsection}{0pt}{*2.5}{2.5pt}
\titlespacing*{\subsubsection}{0pt}{*2}{2pt}

\usepackage[skip=0pt,font=scriptsize]{subcaption}

\usepackage{titling}
\begin{document}

\title{Disc-Cover Complexity Trends in Music Illustrations\\ from Sinatra to Swift }
\author{Nicolas Fracaro$^1$, Stefano Cecconello$^1$, Mauro Conti$^{1,2}$, Niccolò Di Marco$^{3}$, Alessandro Galeazzi$^{1}$ \\[1ex]
$^{1}$University of Padova, Italy,  $^{2}$Örebro University, Sweden,  $^{3}$University of Tuscia, Italy \\[1ex]
nicolas.fracaro@studenti.unipd.it, stefano.cecconello@unipd.it, mauro.conti@unipd.it,\\ niccolo.dimarco@unitus.it, alessandro.galeazzi@unipd.it}
\date{}

\maketitle

\begin{abstract}
The study of art evolution has provided valuable insights into societal change, often revealing long-term patterns of simplification and transformation. Album covers represent a distinctive yet understudied form of visual art that has both shaped and been shaped by cultural, technological, and commercial dynamics over the past century. As highly visible artifacts at the intersection of art and commerce, they offer a unique lens through which to study cultural evolution.

In this work, we examine the visual complexity of album covers spanning 75 years and 11 popular musical genres. Using a diverse set of computational measures that capture multiple dimensions of visual complexity, our analysis reveals a broad shift toward minimalism across most genres, with notable exceptions that highlight the heterogeneity of aesthetic trends. At the same time, we observe growing variance over time, with many covers continuing to display high levels of abstraction and intricacy. Together, these findings position album covers as a rich, quantifiable archive of cultural history and underscore the value of computational approaches in the systematic study of the arts, bridging quantitative analysis with aesthetic and cultural inquiry.
\end{abstract}

\section{Introduction}
Art, in its various forms, has always played a central role in human civilization, shaping cultural identities and reflecting societal changes. Visual arts, including sculptures, paintings and photographs, have been extensively studied, with both qualitative and quantitative analyses revealing characteristic patterns of subject matter and visual complexity across historical periods \cite{sigaki2018history,Taylor1999,PEDRAM2008}. Music, likewise, represents a primary cultural medium, and recent studies have highlighted a general tendency toward simplification in both musical structure and lyrics \cite{DiMarco2025,ParadaCabaleiro2024}.

Over the past century, the global expansion of the music industry and the diversification of distribution formats, from vinyl and cassettes to CDs and digital streaming, have transformed not only how music circulates but also how it is presented to audiences \cite{Brusila2021}. Album covers, initially conceived as purely promotional tools, have gradually acquired cultural and artistic significance. Much like advertisements, they both mirror and shape audience preferences, encapsulating the aesthetics of specific genres and historical contexts~\cite{jones1999steve}.

Beyond their artistic dimension, album covers function as cultural and communicative artifacts, comparable to other visual media. Their evolution reflects changing market strategies and audience expectations, making them a valuable case for quantitative media analysis and for understanding how cultural products visually adapt over time.

In this work, we investigate the evolution of album cover design by analyzing their visual complexity across time and musical genres. Starting from curated collections such as the Billboard Top 200, we build a large-scale dataset of 46.000 album covers spanning 75 years and multiple genres. Following previous research~\cite{sigaki2018history, grebenkina2018edge, valensise2021entropy, wang2024complexity}, we apply information-theoretic and complexity-based measures to quantify their evolution, highlighting both global simplification trends and genre-specific divergences. In particular, while most genres exhibit a reduction in visual complexity, some—such as metal—show the opposite trend. Machine learning–based object recognition further characterizes this divergence, revealing a decreasing number of depicted elements over time. This suggests a tendency toward growing visual minimalism in most genres, whereas others increasingly adopt abstract yet complex imagery, often featuring fused or distorted forms that hinder object recognition. Our study positions album covers as a rich, quantifiable lens on cultural and aesthetic evolution, bridging media analysis and art history, and emphasizes the value of quantitatively examining diverse art forms to achieve a deeper understanding of societal cultural shifts.
\section{Related works}
Our study sits at the intersection of visual complexity analysis, machine learning–based image interpretation, and object recognition, aiming to understand how album covers evolve across time and genres. While previous research has addressed each of these areas individually, few works have combined them to investigate temporal and genre-specific trends in visual artifacts.

\subsection{Visual Complexity Analysis}
The notion of \textit{visual complexity} has been extensively studied across disciplines, but remains without a universally accepted definition\cite{donderi2006visual}. From a computational perspective, Kolmogorov complexity formalizes it as the length of the shortest program that can reproduce an image \cite{li2008introduction}. Since this measure is not computable in practice, compression-based metrics have been developed as practical approximations, estimating Kolmogorov complexity by exploiting the relationship between repetitiveness and compressibility \cite{palumbo2015computerized, marin2013examining}. Alternative approaches for measuring visual complexity include frequency-domain techniques, such as Fourier or Wavelet transforms, which capture the presence of high-frequency components linked to visual detail \cite{redies2007fractal}, and fractal dimension measures that quantify structural irregularity \cite{lam2002evaluation}. More recent methods derive from information theory, including \textit{Minimum Description Length clustering} (MDLc), which measures ``meaningful complexity'' by distinguishing structured signal from random noise \cite{mahon2024minimum}, and the \textit{Entropy--Complexity plane}, which maps images according to their degree of disorder and structural richness \cite{sigaki2018history}.

Beyond computational formalizations, interdisciplinary studies have applied complexity metrics to biology \cite{adami2002complexity}, urban evolution \cite{batty2016complexity}, and cultural products. In visual arts, Sigaki et al.~\cite{sigaki2018history} traced historical trajectories of painting styles using entropy and complexity, while in music, network-based approaches have revealed simplification trends in harmonic and melodic structures \cite{DiMarco2025}. Similarly, analysis of song lyrics has documented the decrease in lexical richness and increased repetitiveness over time \cite{ParadaCabaleiro2024}. These findings support the idea that cultural artifacts undergo systematic simplification processes, motivating similar investigations in album cover design.

\subsection{Machine Learning Approaches to Album Cover Analysis}
Research specifically targeting \textit{album covers} has primarily focused on classification and perceptual studies. Early works used low-level features such as color histograms, textures, and facial presence for genre classification \cite{libeks2011you}. Deep learning approaches, particularly convolutional neural networks (CNNs), have significantly improved classification performance by extracting high-level representations from visual data \cite{oramas2018}. More recently, Greenfield and Paintsil \cite{greenfield2024album} leveraged balanced datasets and modern architectures such as DenseNet-201 and Vision Transformers, outperforming previous CNN-based methods. In addition, studies have highlighted correlations between simple visual features (e.g., average brightness and hue) and musical genres \cite{dorochowicz2019relationship}.

Beyond classification, perceptual studies have examined how cover design influences listeners’ expectations. Venkatesan et al. \cite{venkatesan2022does} showed that typography styles systematically affect perceived music genre: sharp-edged fonts evoke associations with energetic music (e.g., metal), while rounded fonts are linked to softer genres (e.g., pop, classical). Grebenkina et al. \cite{grebenkina2018edge} introduced complexity metrics into this domain, demonstrating that \textit{edge-orientation entropy} correlates positively with human aesthetic preference, including in album covers. This suggests that structural measures can approximate perceptual judgments.

Overall, prior research has either emphasized classification tasks or isolated perceptual aspects, leaving the temporal evolution of cover complexity largely unexplored. Our work addresses this gap by applying information-theoretic and machine learning-based methods to quantify how visual complexity in album covers has evolved across decades and genres.

\subsection{Object Detection in Visual Art}

Advanced computer vision methods have increasingly been applied to the analysis of cultural artifacts, enabling large-scale quantitative studies of visual patterns and trends \cite{manovich2020,lang2021}. Recent work has applied object detection to systematic iconographic studies, enabling quantitative analysis of visual motifs across large art collections \cite{kanstrup2024}. López and Flexer \cite{LopezObj2025} specifically demonstrated zero-shot object detection in album covers for iconographic analysis. Our work extends this approach by using object detection to provide semantic context that enriches the interpretation of visual complexity metrics, enabling a more comprehensive understanding of visual evolution in album cover design.

\section{Methodology}
\label{sec:methods}
In this section we describe the dataset construction and analysis methods chosen for this study. We describe the aggregation and preprocessing of album cover data from multiple sources, and we present three distinct complexity metrics to measure visual complexity and the object detection model used to analyze semantic content.

\subsection{Dataset}
\begin{figure}[h!] 
\begin{subfigure}[t]{\linewidth}
    \centering 
    \includegraphics[width=\columnwidth, alt={}] {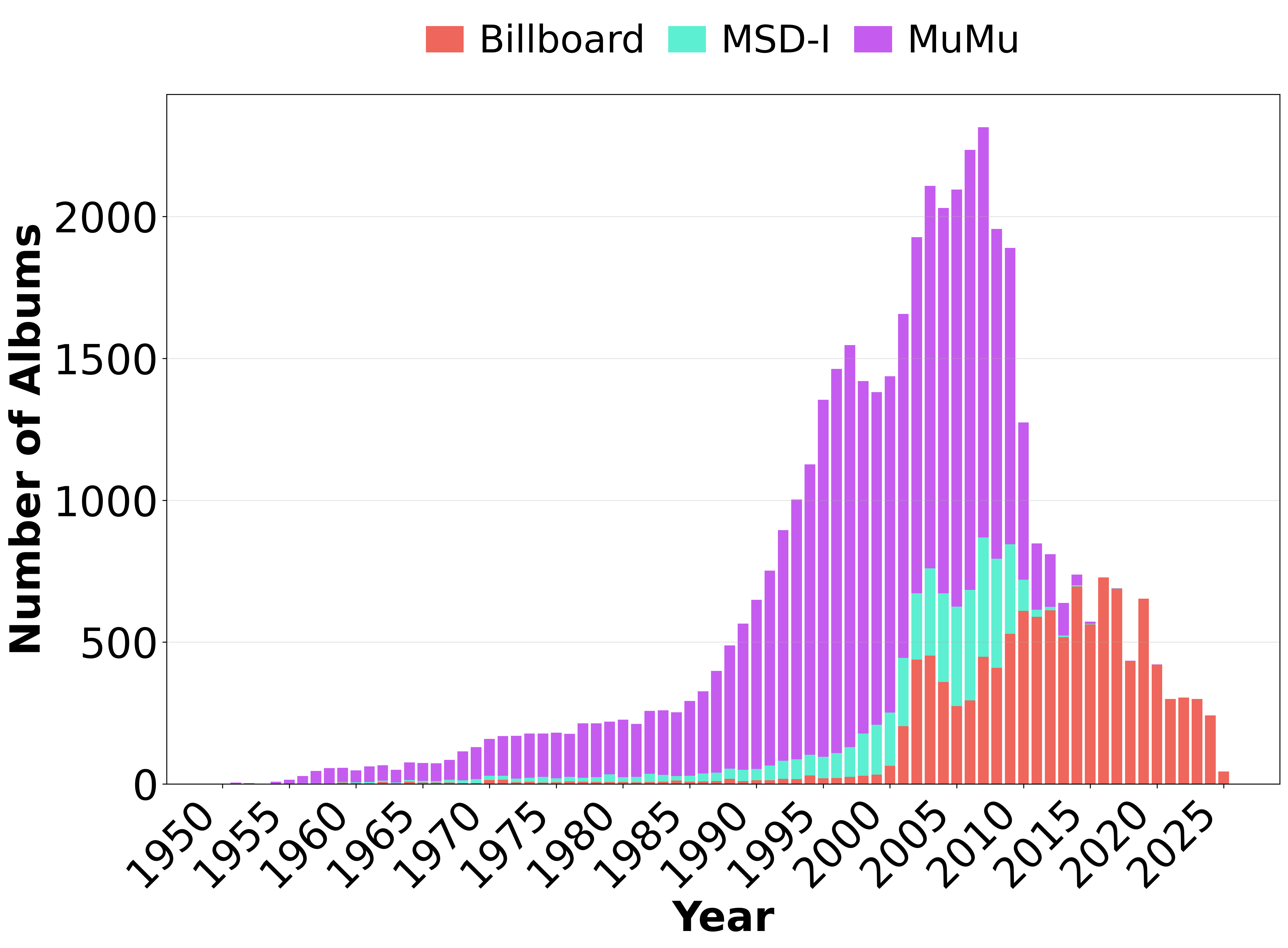}      \caption{Yearly counts of unique albums, colored by retrieval source to highlight the contribution of each dataset.} \label{fig:dataset_composition} 
\end{subfigure}

\begin{subfigure}[t]{\linewidth}
    \centering 
    \includegraphics[width=\columnwidth, alt={Stacked bar chart showing the number of album-genre associations per year, broken down by genre. The chart shows a peak in album counts between 1990 and 2010, with Pop and Rock being the most dominant genres.}] {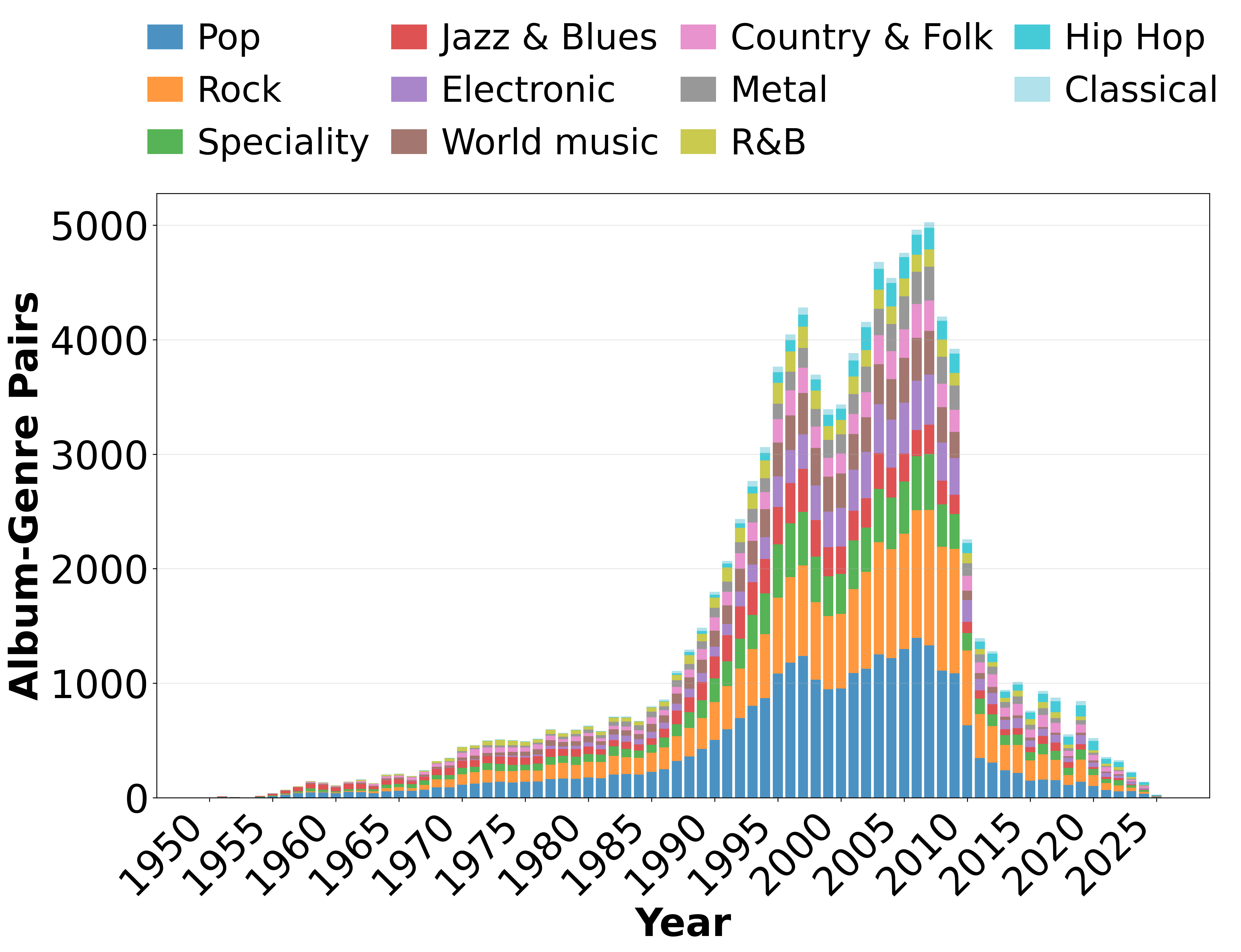}      \caption{Yearly genre distribution of the final dataset. Each bar represents the total number of album-genre associations for a given year, with colored segments indicating the breakdown by genre.} \label{fig:stacked_dataset} 
\end{subfigure}
\end{figure} 

\paragraph{Data Sources and Aggregation} The dataset for this study was constructed by aggregating and standardizing data from three distinct sources. The initial foundation was built upon two existing academic datasets: the MuMu dataset\footnote{\url{https://www.upf.edu/web/mtg/mumu}}, which provided both single- and multi-labeled genre annotations, and the MSD-I dataset\footnote{\url{https://zenodo.org/records/1240485}}, which contained single-labeled albums. To supplement these sources, particularly with more recent releases, we compiled a third dataset by scraping the Billboard charts\footnote{\url{https://www.billboard.com/charts/}} from January 2002 to April 2025. This was accomplished using the \texttt{billboard.py} library to programmatically access twenty chart categories\footnote{The scraped Billboard categories include: comedy, world, new-age, reggae, christian, tropical, traditional-jazz, rock, regional-mexican, latin-rhythm, latin-pop, independent, compilation, classical-crossover, classical, cast, americana-folk, billboard-200, vinyl, and tastemaker.}. For all sources, the MusicBrainz database\footnote{\url{https://musicbrainz.org/}} was used as the canonical reference to standardize album metadata, ensuring that for each entry, we retrieved the cover art corresponding to its first official release.

\paragraph{Deduplication and Cleaning} A multi-stage deduplication and cleaning process was implemented to ensure the uniqueness of each album. Initially, duplicates between the MuMu and MSD-I datasets were removed by cross-referencing their respective Million Song Dataset and MusicBrainz identifiers. Subsequently, albums from the Billboard source were deduplicated against the existing set using artist and title pairs. The final deduplication step was performed once all albums were mapped to a MusicBrainz ID for their first release, using this identifier as the primary key. For the cleaning phase, albums with broken image URLs or corrupted files that could not be processed by our complexity algorithms were also removed. During this process, 5,096 albums from the Billboard dataset were excluded due to faulty or missing URLs that prevented image download, while 20 images were identified as corrupted and could not be processed by the complexity estimation algorithms.

\paragraph{Genre Standardization} The raw genre labels aggregated from the different sources were highly heterogeneous, comprising over 1,000 unique strings that included specific sub-genres, non-genre tags (e.g., "acoustic", "1990"), and other metadata not relevant for our study. To create a consistent and meaningful set of categories for analysis, we first defined a target set of 11 primary genres \footnote{The genres include: Pop, Rock, Speciality, Jazz \& Blues, Electronic, World Music, Country \& Folk, Metal, R\&B, Hip Hop, Classical}, guided by the taxonomy provided by Musicmap \footnote{Musicmap. The Genealogy and History of Popular Music Genres. \url{https://musicmap.info}. Accessed: July 2025.}. We then employed the Gemini 2.5 Pro Large Language Model via its public web-based chat interface to map each of the 1,000+ raw labels to one or more of these 11 target genres, as well as to identify and discard non-genre-related tags. This approach was chosen because the number of unique labels to be mapped was relatively small, making it feasible to complete within the interface's free usage limits.

\paragraph{LLM-based Data Imputation} A subset of the albums, primarily from the Billboard source, lacked genre information entirely. To address this, we performed automated genre prediction for the 2,510 unlabeled albums programmatically via the Google AI Python API. We employed the `gemini-2.0-flash` model for this task, as it was suited for the large-scale batch processing required for our analysis and was the freely accessible model via the API at the time of the research. We designed a structured prompt that constrained the model to classify albums using only our predefined 11 supergenres and return results in a standardized CSV format with confidence indicators \footnote{Full prompt: "You are a music expert tasked with classifying album genres. I will provide you with a list of albums, and you need to classify each one using ONLY the following 11 supergenres: Classical,Country \& Folk, Electronic, Hip Hop, Jazz \& Blues, Metal, Pop, R\&B, Rock, Speciality, World music
IMPORTANT INSTRUCTIONS:
1. Return ONLY a CSV format with the following columns: album\_group\_mbid, title, release\_group\_date, genres, sure
2. The genres column should contain a Python list format: ['Genre'], only 1 single genre per album is allowed
3. Use ONLY the 11 supergenres listed above - no other genres are allowed
4. If you're unsure about a genre, pick the most likely one from the 11 options but set "sure" to false
5. Do not include any explanations, headers, or additional text - ONLY the CSV data
6. Do not include markdown formatting or code blocks
7. CRITICAL: Use proper CSV escaping - put double quotes around any field that contains commas, quotes, or special characters
8. Example format: "album\_id","Title, with comma","2023","['Pop']","true"
Here are the albums to classify:}. 

To validate this approach, we tested the model on a control set of 500 albums with known labels, instructing it to predict a single genre from our set of 11 and to provide a boolean confidence value. The model reported high confidence for 450 predictions, of which 87\% were correctly contained within the album's true genre labels. For the 50 low-confidence predictions, this accuracy dropped to 62\%. Based on these results, we proceeded with the imputation but only accepted high-confidence predictions. Finally, 911 albums in the dataset were missing a release date. We explored using the `gemini-2.0-flash` model to predict the year of release. However, a validation test on 500 albums with known dates yielded unsatisfactory results. The model returned high-confidence predictions for only 289 cases. For this subset, the accuracy for predicting the exact year was 64.4\%, rising to 80.3\% for a tolerance of $\pm$5 years. Given the low confidence rate and insufficient accuracy, we deemed this approach unreliable for our time-series analysis. Consequently, we opted to remove these 911 albums from the dataset to maintain the temporal integrity of our analysis.

\paragraph{Final Dataset Characteristics} After the processes of cleaning, deduplication, and standardization, the final dataset consists of 46,399 unique, multi-labeled albums, each with a corresponding cover image. 
The dataset composition is presented in Figure~\ref{fig:dataset_composition}, which reports the number of unique albums per year by source. On the other hand, Figure~\ref{fig:stacked_dataset} illustrates the genre distribution over time, with albums associated with multiple genres counted more than once.
It is important to note the presence of two main biases: a genre bias, with a predominance of Pop and Rock albums, and a temporal bias, with the majority of albums concentrated between 1990 and 2010. Furthermore, as a multi-labeled dataset, a single album can be associated with, and thus contribute to, more than one genre category. 
A specific feature of the temporal distribution is the decrease in the number of albums between 1999 and 2003. This trend was already present in the publicly available MuMu and MSD-I datasets, and became slightly more pronounced when extending coverage with Billboard data starting from 2002. 
Despite these biases, our final dataset provides a reasonable representation of the musical landscape, encompassing the most popular genres over a span of 75 years.

\subsection{Complexity Metrics} 
\label{sec:complexity_metrics}
To conduct a robust and multifaceted analysis, we adopted a multi-metric approach to quantify the visual complexity of album covers. As no single metric can capture all aspects of complexity, we selected distinct computational methods that provide a comprehensive overview, complemented by an object detection analysis to add a semantic dimension.

\paragraph{Entropy-Complexity Plane} To disentangle visual disorder from structural richness, we utilized the Entropy-Complexity plane methodology developed by Sigaki et al.~\cite{sigaki2018history}. This approach was chosen for its ability to characterize an image using two complementary measures, both bounded between 0 and 1: Permutation Entropy (H), which quantifies the degree of randomness in local pixel patterns, and Statistical Complexity (C), which measures the presence of elaborate spatial structures. An image is thus represented as a point in a 2D plane, allowing for a more nuanced analysis than a single complexity score. For example, an image of random noise would exhibit high entropy (approaching 1) due to its unpredictable pixel patterns, but low statistical complexity (near 0) because it lacks coherent spatial structures. A solid white image would share this low statistical complexity due to its lack of spatial structure, while showing near-zero entropy because of its complete predictability. As a pre-processing step, each image was converted to its grayscale representation. The calculation is then based on the distribution of local ordinal patterns (2x2 pixel windows in our implementation), as described in the original paper. To illustrate the meaning of the two measures in some real-case examples, Figure \ref{fig:four-images} presents the covers of four famous musical albums, positioned according to their values of statistical complexity and permutation entropy.

\paragraph{Meaningful Complexity (MDLc)} To measure the ``meaningful" complexity of an image, we implemented the method proposed by Mahon and Lukasiewicz~\cite{mahon2024minimum}. MDLc values range from 0 (indicating minimal meaningful structure) to theoretically unbounded positive values, where higher scores represent more complex images. We selected this metric for its specific design to distinguish structured, significant information from random noise. Unlike traditional metrics that might assign high complexity to a textureless, noisy image, MDLc uses a hierarchical clustering approach based on the Minimum Description Length (MDL) principle to identify and quantify coherent structures at multiple scales. For our implementation, we used the authors' publicly available code\footnote{\url{https://github.com/Lou1sM/meaningful_image_complexity?tab=readme-ov-file}}. All images were resized to a standard resolution of 224x224 pixels to ensure consistency. To make the computation feasible on our large dataset, we reduced the maximum number of clusters ($K_{max}$) to test during the MDL search from 8 to 5, a modification suggested by the original authors to improve efficiency without significantly impacting accuracy. All other hyperparameters were kept at their default values.

\paragraph{Compression-based Complexity (ZIPc)} As a simple and objective baseline, we included a compression-based metric, ZIPc. This approach provides a practical approximation of Kolmogorov complexity, where the complexity of an image is estimated by its incompressibility. We chose this metric because it offers a robust measure of statistical redundancy that is straightforward to compute and interpret. The ZIPc score was calculated as the ratio of the compressed file size to the original raw bitmap size. 
The metric typically produces values between 0 and 1, where 0 represents perfect compressibility (minimal complexity) and values approaching 1 indicate high incompressibility (maximum complexity). In rare cases, values may exceed 1 when compression overhead surpasses any achieved compression gains.
To ensure a standardized process, each image was first converted to an uncompressed RGB bitmap. The bitmap data was then compressed using the DEFLATE algorithm (via the ZIP file format) with the compression level set to its maximum value (9).
\paragraph{Semantic Complexity via Object Detection} To complement the perceptual complexity metrics, we introduced a measure of semantic complexity based on the number of recognizable objects in an image. We employed the YOLOv8 model~\cite{yolov8}, specifically the lightweight yolov8n.pt version pre-trained on the COCO dataset. This model was chosen for its high efficiency and accuracy in identifying a predefined set of 80 common object classes. For each album cover, we extracted both the total count of detected objects and their corresponding object classes using a confidence threshold of 0.25. Object counts range from 0 (no detected objects) to theoretically unbounded positive integers. This approach provides both a quantitative measure of the "busyness" of album covers in terms of recognizable content and qualitative insights into the types of objects commonly featured across different genres.
\section{Results and Discussion}
This section presents and discusses the results of the analysis of temporal evolution and genre differences in album cover complexity using three quantitative metrics, supplemented by automated object detection to identify the semantic drivers underlying the observed patterns.

\subsection{Entropy-Complexity Plane Metric}
\begin{figure}[htbp]
    \centering

    \begin{tabular}{ccc}
        \raisebox{0.5cm}{\rotatebox{90}{\textbf{High S. Complexity}}} & 
        \begin{subfigure}{0.35\columnwidth}
            \includegraphics[width=\textwidth]{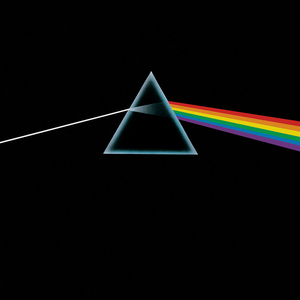}
            \label{fig:img_a}
        \end{subfigure} & 
        \begin{subfigure}{0.35\columnwidth}
            \includegraphics[width=\textwidth]{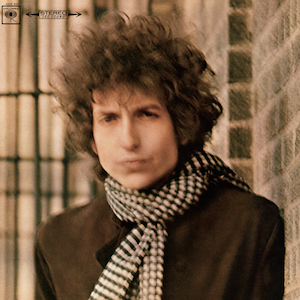}
            \label{fig:img_b}
        \end{subfigure} \\[0.5cm]
        
        \raisebox{0.5cm}{\rotatebox{90}{\textbf{Low S. Complexity}}} & 
        \begin{subfigure}{0.35\columnwidth}
            \includegraphics[width=\textwidth, trim={5cm 5cm 5cm 5cm},clip]{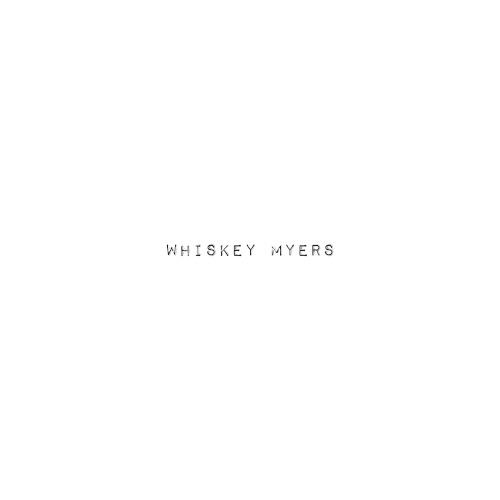}
            \label{fig:img_c}
        \end{subfigure} & 
        \begin{subfigure}{0.35\columnwidth}
            \includegraphics[width=\textwidth]{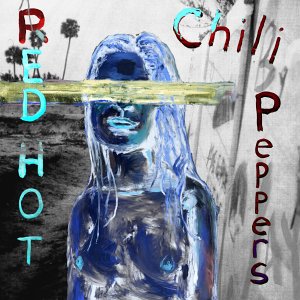}
            \label{fig:img_d}
        \end{subfigure} \\
        & \textbf{Low Entropy} & \textbf{High Entropy}
    \end{tabular}
    \caption{Album cover examples with different complexity and entropy scores. (top left) Pink Floyd, The Dark Side of the Moon, 1973. (top right) Bob Dylan, Blonde on Blonde, 1966. (c) Whiskey Myers, Whiskey Myers, 2019. (d) By the Way, Red Hot Chili Peppers, 2002.}
    \label{fig:four-images}
\end{figure}

We begin our analysis by considering the entropy–complexity plane. As discussed in Section~\ref{sec:methods}, this representation provides a nuanced perspective on the visual characteristics of album covers, revealing distinct patterns across genres and over time. To build intuition, Figure~\ref{fig:four-images} presents four representative examples that illustrate the visual properties associated with different regions of the plane. The two columns correspond to low (left) and high (right) entropy, while the two rows represent low (bottom) and high (top) complexity.
Low-entropy covers are generally characterized by uniform backgrounds and a small number of orderly visual elements, in sharp contrast to the random-like noise associated with high entropy. Within this group, the Whiskey Myers album (bottom left) is visually simple, consisting only of a text element, whereas the Pink Floyd album (top left) demonstrates higher complexity through highly regular patterns, such as the triangular shape and colored stripes.
In contrast, high-entropy covers are marked by more chaotic, blurred contours and shaded elements, which increase their resemblance to random noise. In this region, background structure plays a decisive role in differentiating levels of complexity. For example, the Bob Dylan album (top right) exhibits higher statistical complexity due to structured patterns such as the brick wall and scarf, while the Red Hot Chili Peppers album (bottom right) lacks comparable regularity and therefore presents lower complexity despite its high entropy.

\begin{figure}[ht]
\centering
\begin{subfigure}[t]{\linewidth}
\centering
\includegraphics[width=\columnwidth,
alt={Scatter plot showing mean permutation entropy versus statistical complexity for 11 music genres. Most genres cluster in the center, with Electronic, Hip Hop, and Metal as distinct outliers.}]
{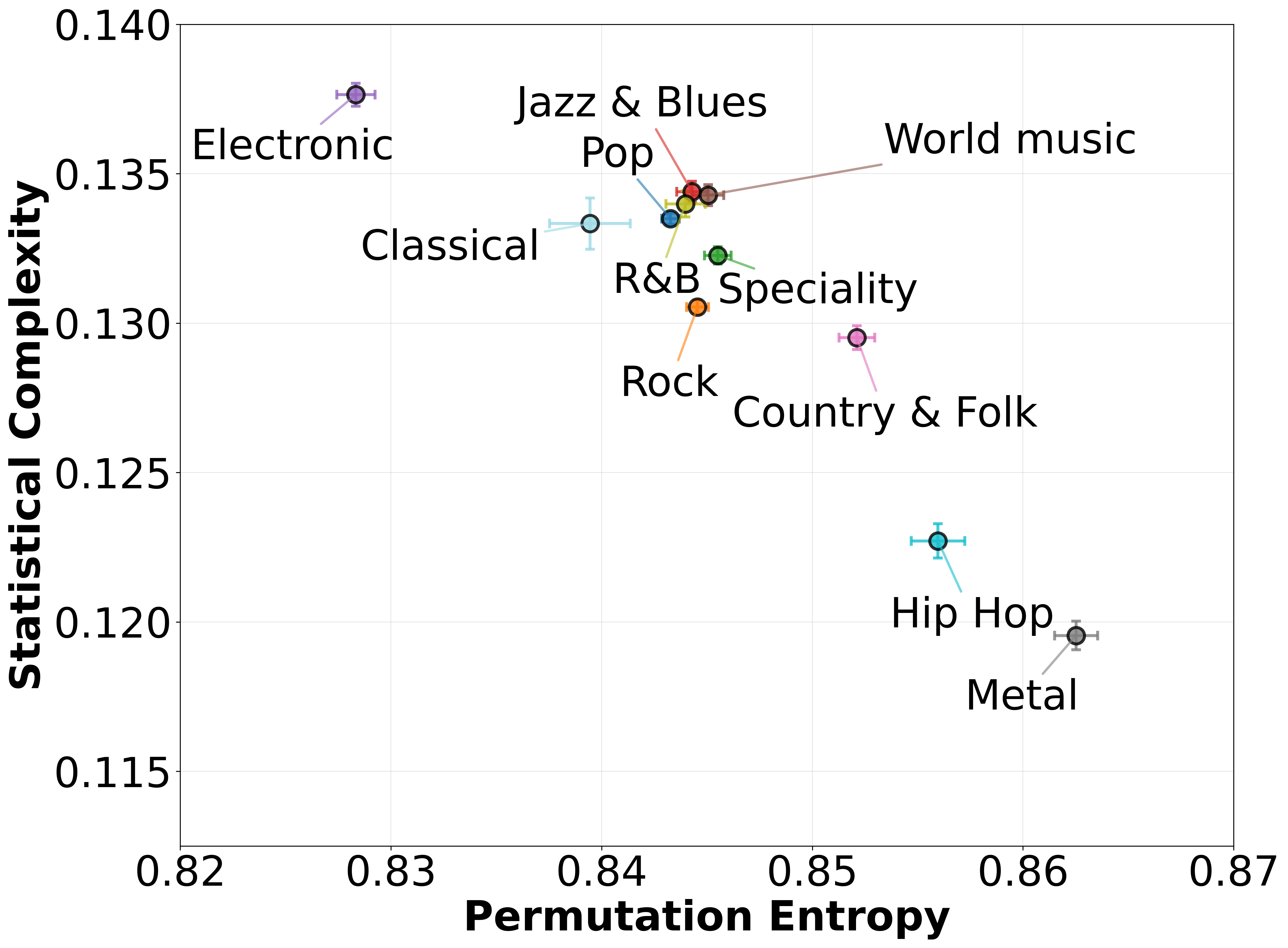}
\caption{Mean Permutation Entropy and Statistical Complexity by genre.}
\label{fig:entropy_complexity_by_genre}
\end{subfigure}
\hfill
\begin{subfigure}[htp]{\linewidth}
\centering
\includegraphics[width=\columnwidth,
alt={A plot showing the trajectory of average entropy and complexity values over chronological time periods from 1950 to 2025. The path shows a distinct shift towards lower statistical complexity in recent periods.}]
{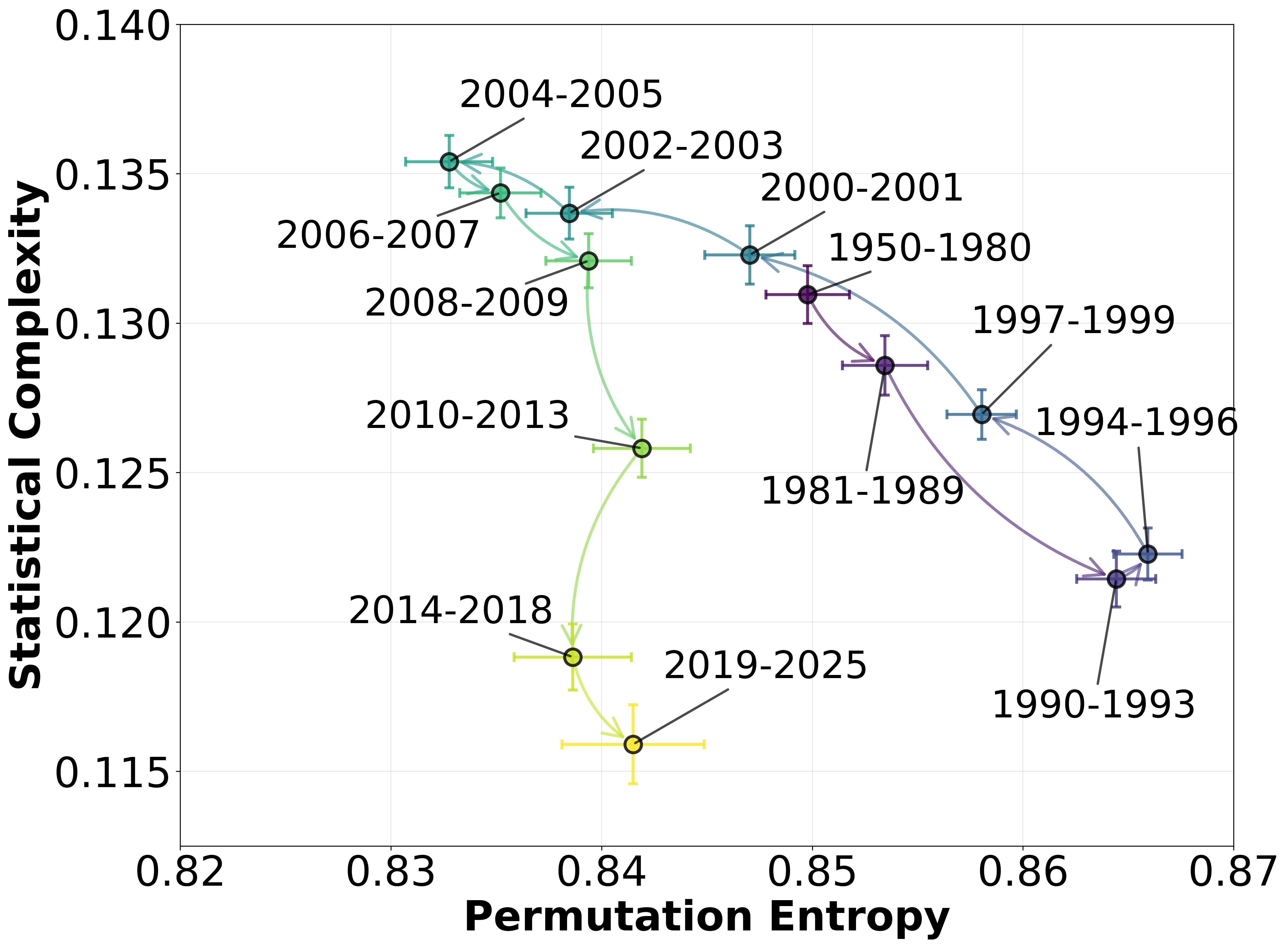}
\caption{Evolutionary trajectory of mean complexity values over time. Periods were defined of variable duration to ensure a minimum of 3000 albums per period (except the final period, which contains 2,269)}
\label{fig:entropy_complexity_trajectory}
\end{subfigure}
\caption{Characterization of album covers in the Entropy-Complexity plane. Panel \textit{(a)} shows the average positioning for each music genre, highlighting a central cluster and key outliers. Panel \textit{(b)} illustrates the aggregate temporal evolution across all genres.}
\label{fig:ec_plane_combined}
\end{figure}

Figure~\ref{fig:entropy_complexity_by_genre} shows the Entropy-Complexity plane grouped by genres, where each point represents the mean entropy and complexity values for that genre. Interestingly, the figure reveals a dense central cluster comprising a majority of the music genres. The overlapping standard error bars confirm the statistical significance of this grouping, which points to a widely adopted stylistic norm of moderate permutation entropy and statistical complexity. Beyond this central cluster, three genres emerge as distinct outliers. Metal and Hip Hop are characterized by higher entropy and lower statistical complexity, a profile that may reflect more visually disordered aesthetics. Conversely, the Electronic genre occupies an opposing region of lower entropy and higher statistical complexity, which suggests more structured, minimalist designs.

We repeat the analysis grouping albums over time. The results are shown in Figure~\ref{fig:entropy_complexity_trajectory}, where time is divided into dynamic periods containing approximately 3000 albums each. This threshold balances data sufficiency with temporal granularity, ensuring each period contains adequate data for stable comparisons while preserving the ability to detect temporal trends. The results reveal a clear, non-random evolutionary trajectory that can be grouped into four macro-periods. Initially, from the 1950s to 1996, a rise in permutation entropy and a fall in statistical complexity suggest a trend toward more disordered artwork. This pattern shifted between 1997 and 2003, as lower entropy and higher statistical complexity indicate a shift toward more structured designs. A period of relative stability followed from 2004 to 2009. Finally, the most recent period (2010–2025) is marked by a sharp decrease of approximately 12\% in statistical complexity at stable entropy levels. Since entropy remains largely unchanged, indicating consistent visual disorder, the observed decline in statistical complexity reflects a shift toward compositions with reduced structural elaboration. This trajectory points to a move toward more streamlined designs, with fewer intricate spatial patterns in the modern era.

Several factors may be at the core of the observed trend, encompassing both cultural shifts and technological developments. Culturally, the rising popularity of metal and hip hop during the 1980s and 1990s may have influenced the visual style of album covers in other genres, contributing to increased permutation entropy alongside reduced statistical complexity during this period. From a technological perspective, the emergence of music-oriented digital platforms, such as iTunes (launched globally in 2004), may have driven the decrease in statistical complexity from 2004 onward. Indeed, these platforms typically displayed album artwork as small thumbnails—by 2010, limited to 600×600 pixels at 72 dpi~\cite{mcveigh2010itunesart}—potentially discouraging the inclusion of highly detailed designs.

\subsection{MDLc and ZIPc Complexity Metrics}
\begin{figure}[ht]
\centering
\begin{subfigure}[t]{\linewidth}
\centering
\includegraphics[width=\linewidth,
alt={Line chart showing the average MDLC complexity score for 11 music genres from 1950 to 2025. Metal and Hip Hop are consistently highest, while Electronic is lowest. Most lines trend downward after 2009.}]
{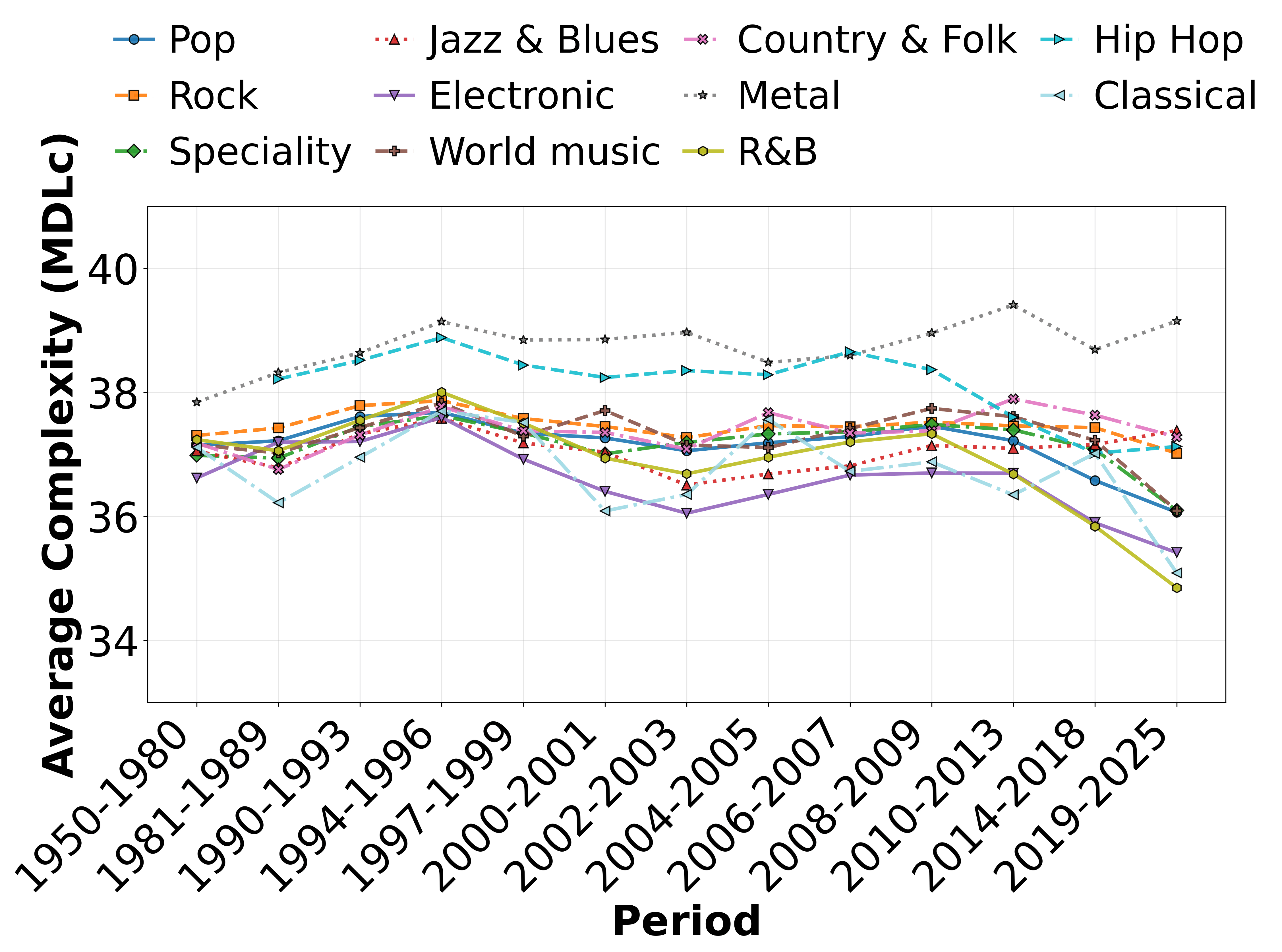}
\caption{Average MDLc score over time.}
\label{fig:mdlc_over_time}
\end{subfigure}
\hfill
\begin{subfigure}[t]{\linewidth}
\centering
\includegraphics[width=\linewidth,
alt={Line chart showing the average ZIPc complexity score for 11 music genres from 1950 to 2025. Metal and Hip Hop are consistently highest, while Electronic is lowest. Most lines trend downward after 2009.}]
{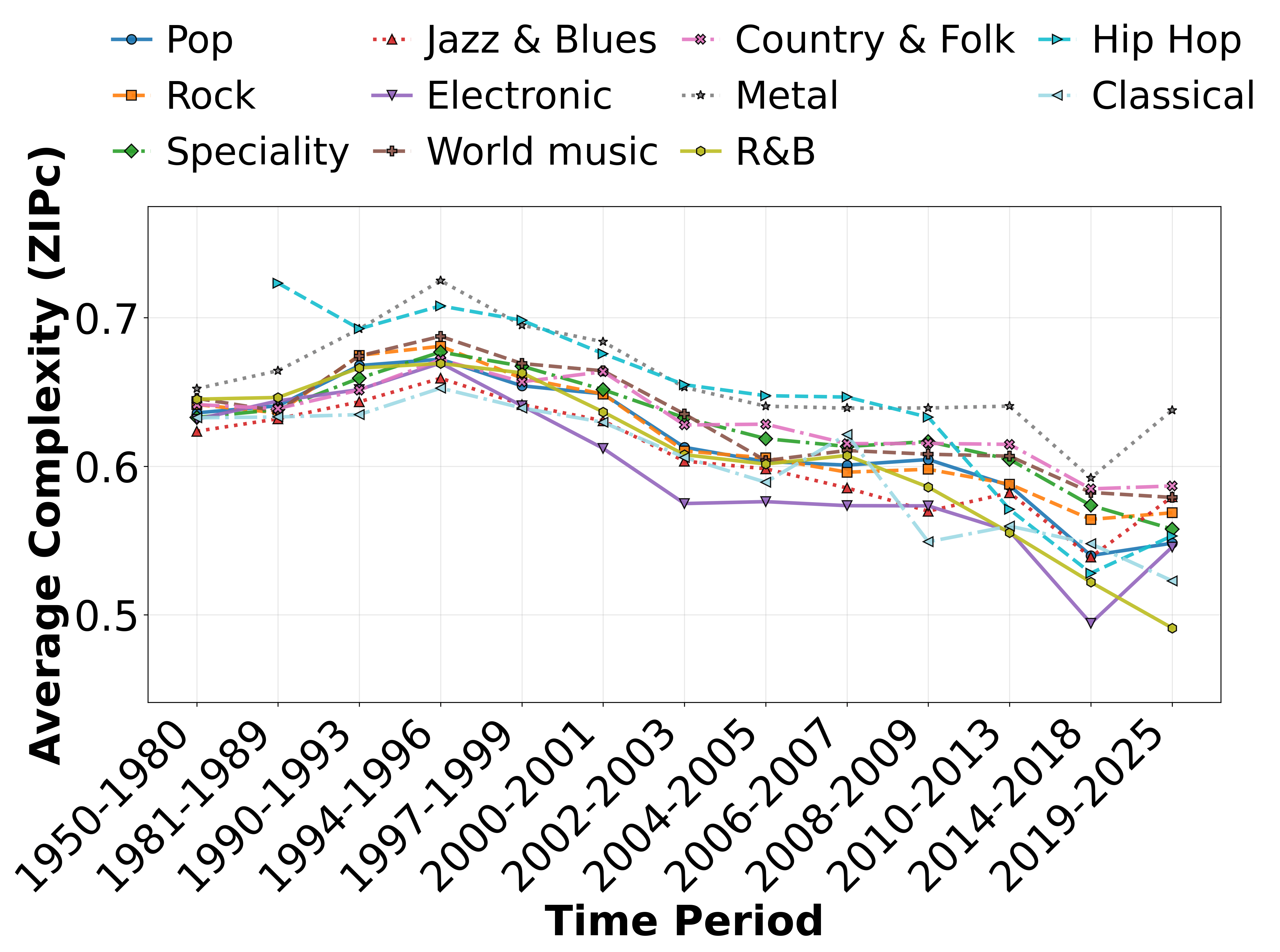}
\caption{Average ZIPc score over time.}
\label{fig:zipc_over_time}
\end{subfigure}
\caption{Temporal evolution of average visual complexity across genres, measured by \textit{(a)} MDLc and \textit{(b)} ZIPc. Periods were defined of variable duration to ensure a minimum of 3000 albums per period (except the final period, which contains 2,269). A genre is included in a period only if at least 50 albums of that genre are present.}
\label{fig:complexity_evolution}
\end{figure}

Analysis of the MDLc, which measures meaningful information by distinguishing coherent patterns from random noise, and ZIPc, which uses compression ratios as an estimate of image complexity, corroborates the findings from the entropy-complexity plane, confirming the notable similarity in visual complexity across most genres, as shown in Figure~\ref{fig:complexity_evolution}.
These metrics show that, while most genres maintain proximate complexity scores throughout the observed timeline, the three previously identified outliers are consistent across these measures: Metal and Hip Hop consistently exhibit the highest average complexity scores, while the Electronic genre consistently registers among the lowest. These results reinforce the characterization of Metal and Hip Hop's aesthetics as visually dense, and Electronic's as minimalist.

The temporal evolution of the MDLc and ZIPc scores also aligns with the four macro-periods identified previously. A general increase in complexity is observed until the mid-1990s, followed by a reversal and a period of stability. Critically, both metrics confirm the significant, widespread decrease in average complexity during the most recent period (2010-2025), lending further support to the hypothesis of a shift towards less elaborate cover art. This finding confirms that the general trend towards simplification, previously documented for lyrical content~\cite{ParadaCabaleiro2024, Varnum2021} and melodic structure~\cite{DiMarco2025}, also extends to the visual dimension of music. This convergence towards simplicity across these domains suggests a shared response to the modern media landscape, where a plausible driver is the shift to digital platforms. In a context where immediate engagement is critical, simpler, bolder designs may be favored for their ability to capture attention quickly, particularly on small screens or in recommendation-driven listening environments, just as more repetitive lyrics and less complex melodies serve similar functions in the auditory domain.

\begin{figure}[ht]
\centering
\begin{subfigure}[t]{0.97\linewidth}
\centering
\includegraphics[width=0.97\linewidth,
alt={Series of box plots for MDLc complexity scores from 1950 to 2025, showing that the boxes become taller in recent years, indicating wider data dispersion.}]
{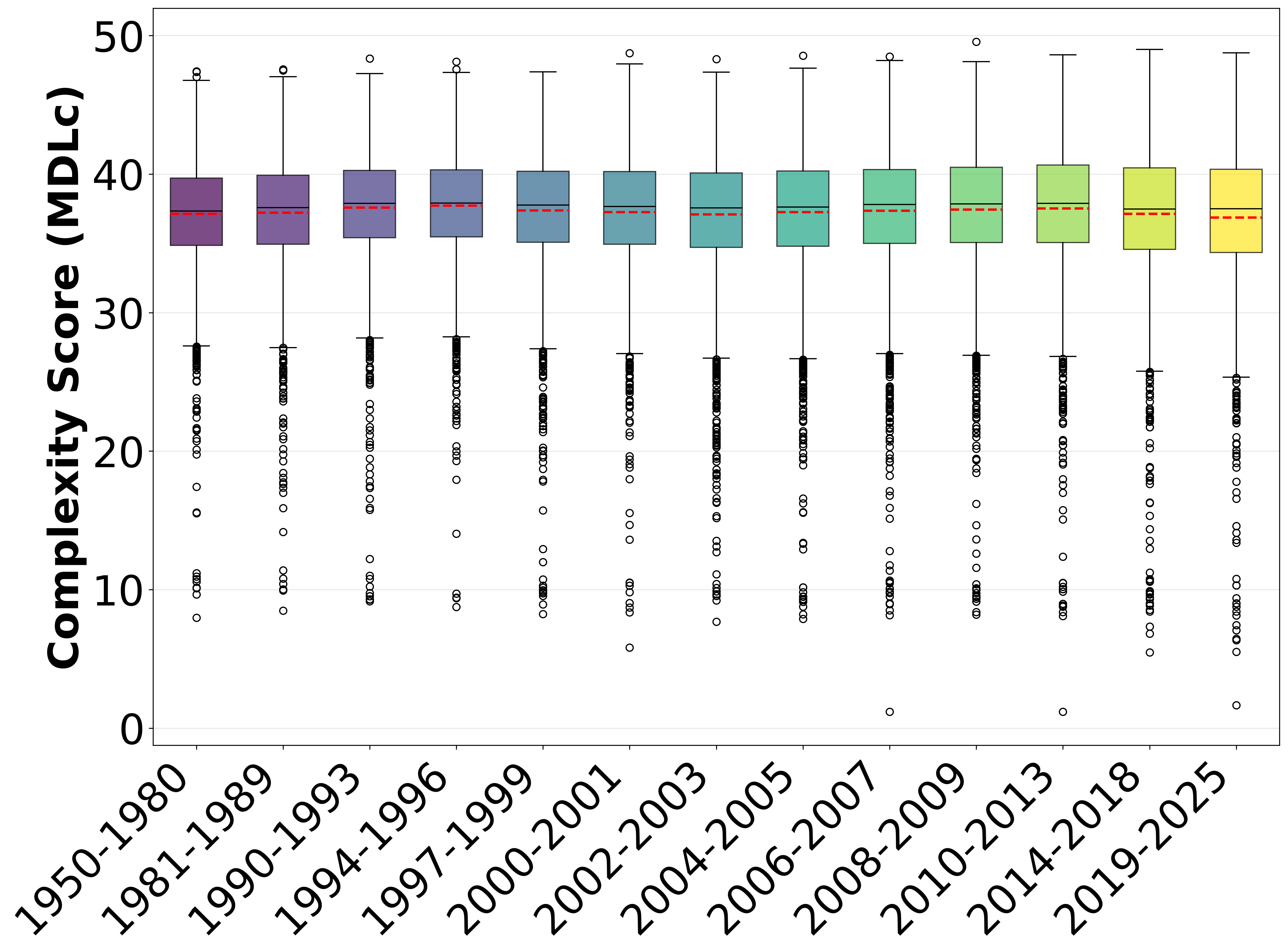}
\caption{Distribution of MDLc scores per period.}
\label{fig:mdlc_distribution_over_time}
\end{subfigure}
\hfill
\begin{subfigure}[t]{0.97\linewidth}
\centering
\includegraphics[width=0.97\linewidth,
alt={Series of box plots for ZIPc complexity scores from 1950 to 2025, showing that the boxes become taller in recent years, indicating wider data dispersion.}]
{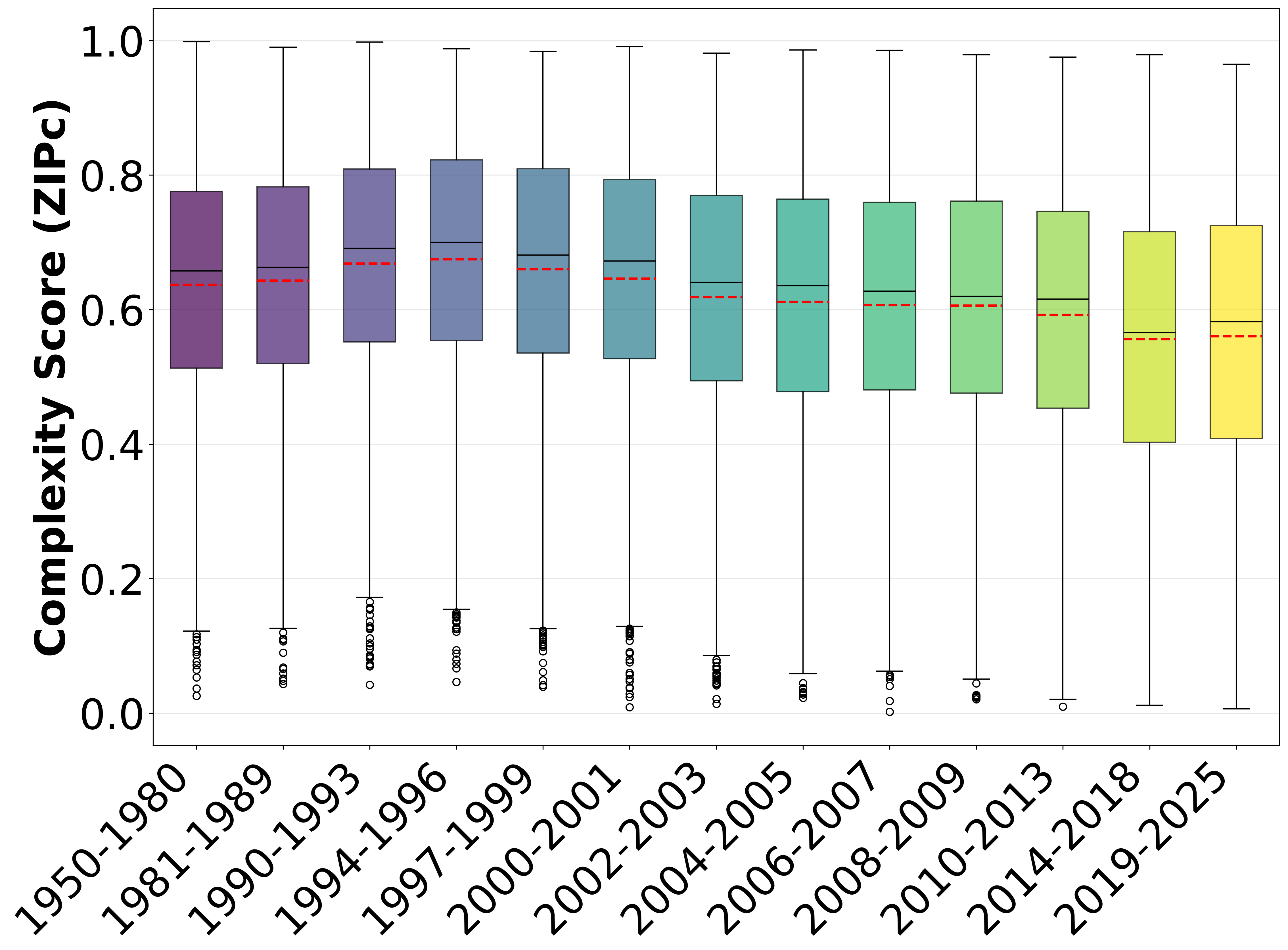}
\caption{Distribution of ZIPc scores per period.}
\label{fig:zipc_distribution_over_time}
\end{subfigure}
\caption{Evolution of the distribution of complexity scores over time for \textit{(a)} the MDLc metric and \textit{(b)} the ZIPc metric. While the median complexity (red dashed line) trends downward in recent years, the interquartile range (height of the boxes) increases, indicating growing stylistic diversity.}
\label{fig:complexity_evolution_distribution}
\end{figure}

However, this trend toward simplification is nuanced by a simultaneous increase in stylistic diversity. While the median complexity has decreased, the dispersion of complexity scores has progressively widened over time, within each period comprising at least 3000 albums, as illustrated in Figure~\ref{fig:complexity_evolution_distribution}. This is most pronounced in the last 15 years, during which the interquartile range has increased by approximately 11\% for both the MDLc and ZIPc metrics. This suggests that the modern media landscape has not led to visual homogenization. Instead, it appears to foster a polarization of aesthetics, where a dominant trend toward minimalist, high-impact designs coexists with the continued exploration of highly complex, niche visual styles. It is plausible that the digital landscape itself creates favorable conditions for this diversity; the democratization of advanced creative software facilitates the production of intricate visuals, while the lowered barrier to entry on digital platforms may reduce the commercial pressures of traditional record labels, potentially creating more space for such artistic experimentation.
Interestingly, most of the outliers appear on the lower end of the distribution. One possible interpretation is that this asymmetry may reflect an intrinsic characteristic of album cover design: while there might be aesthetic constraints on how far visual complexity can be pushed, it may be comparatively easier for designs to deviate toward the minimalist extreme. 

\subsection{Interpreting Complexity through Object Detection}
In the previous section, we applied several complexity metrics to identify overall patterns in album covers, without further elaborating on the underlying shifts of the observed behavior.
To fill this gap, in this section we focus on the semantic content of the covers using the YOLOv8 object detection model to extract both the number and classes of detected objects. 
By applying object detection on album cover, we analyze the trend on the presence and classes over time and across genres.

\begin{figure}[ht!]
\centering
\begin{subfigure}[t]{\linewidth}
\centering
\includegraphics[width=\columnwidth,
alt={Stacked bar chart showing the proportional distribution of the top 10 detected objects for each music genre. The 'person' class is visually dominant in every bar.}]
{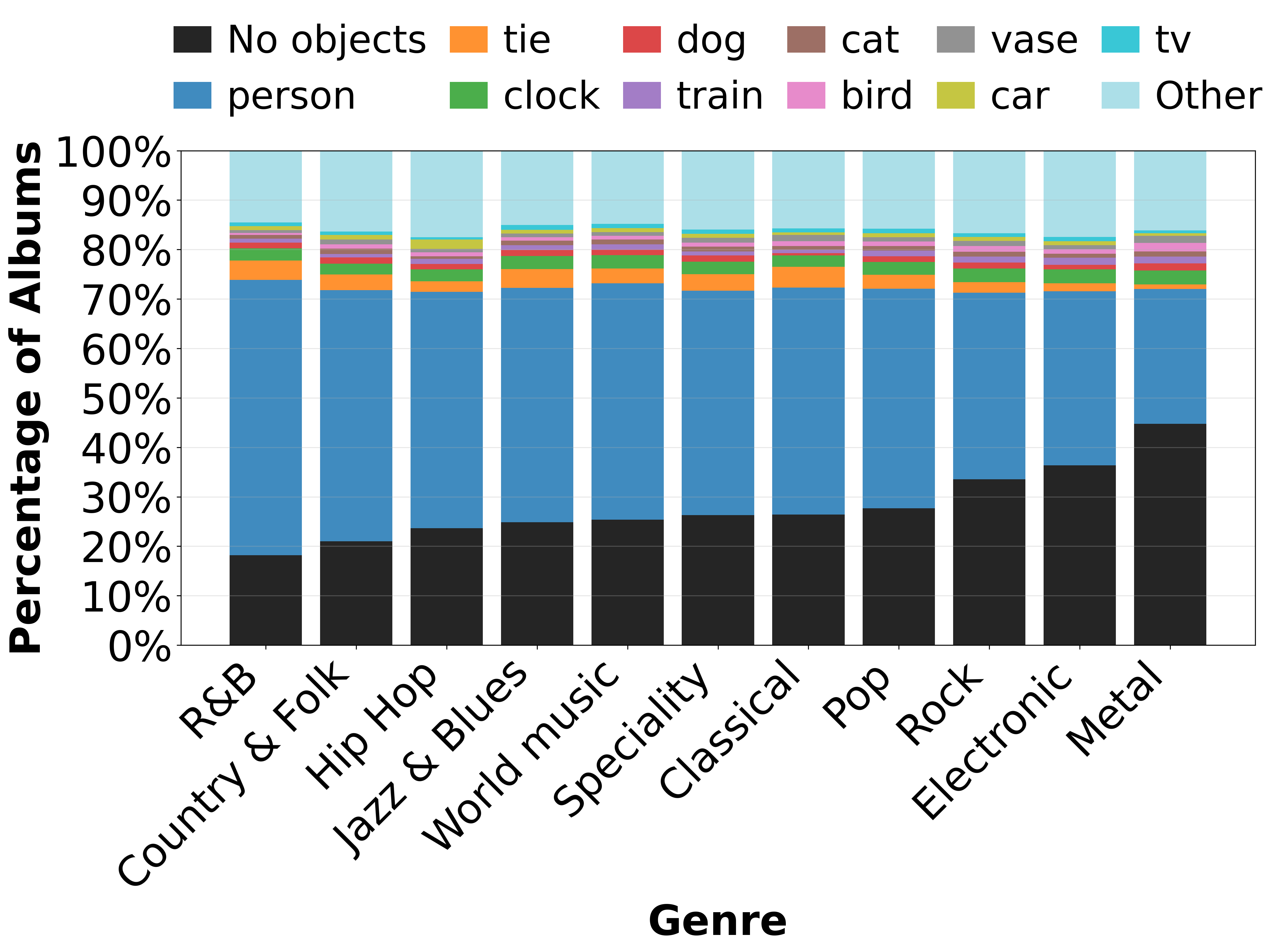}
\caption{Proportional distribution of detected object classes per genre, including albums with no detected objects and the top 10 most frequently detected object classes.}
\label{fig:objects_stacked}
\end{subfigure}
\hfill
\begin{subfigure}[t]{\linewidth}
\centering
\includegraphics[width=\columnwidth,
alt={Line chart tracking the average number of detected objects per album from 1950 to 2025 for 11 genres. Metal consistently shows the lowest line, while Electronic is also among the lowest.}]
{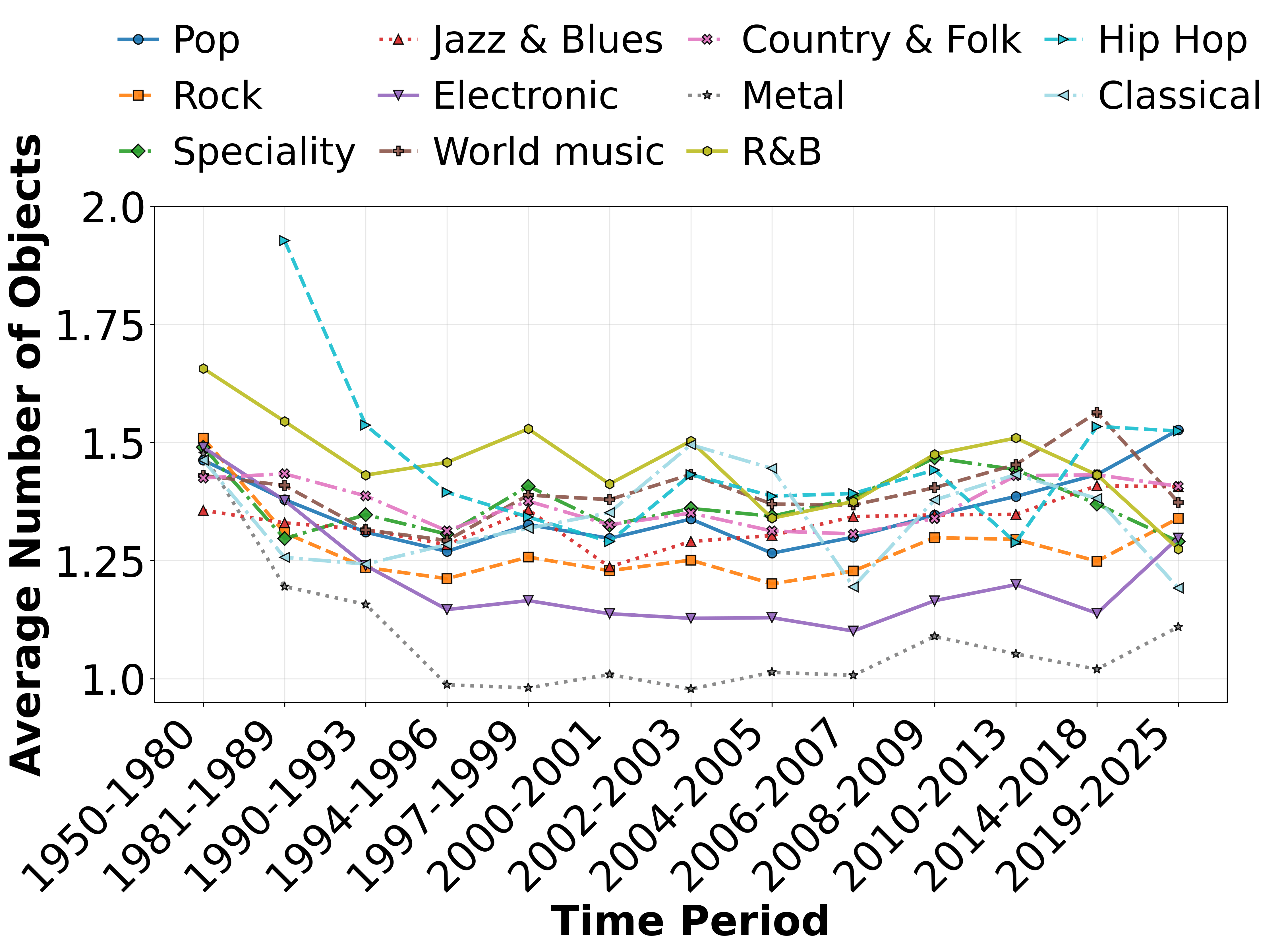}
\caption{Average number of detected objects per album over time.}
\label{fig:objects_over_time}
\end{subfigure}
\caption{Analysis of semantic content via object detection across music genres. Panel \textit{(a)} shows the proportional distribution of detected object classes per genre, including albums with no detected objects. The `person' class predominates across all genres, while Electronic and Metal show higher proportions of albums with no detected objects. Panel \textit{(b)} illustrates the temporal evolution of average object counts per album. Metal consistently exhibits the lowest object count despite high visual complexity, while Electronic maintains low counts consistent with its minimalist aesthetic.}
\label{fig:object_detection}
\end{figure}

As shown in Figure~\ref{fig:objects_stacked}, the ``person" class is the most frequent semantic element across all genres. This widespread use of portraiture, which could serve as a direct marketing tool by placing the artist at the forefront, may help explain the moderate, tightly-grouped complexity scores that form the stylistic norm, potentially reflecting constraints on visual language diversity.

Using the same dynamic periods of at least 3000 albums each, the temporal evolution of object counts (Figure~\ref{fig:objects_over_time}) further clarifies the nature of the outlier genres. The case of Metal is particularly insightful: despite its high complexity, our object detection analysis reveals it has the lowest average number of detected objects. This apparent paradox may indicate that Metal's complexity derives from sources other than recognizable objects, potentially abstract compositions, intricate logos, and detailed textures. In contrast, the low object count for the Electronic genre aligns with its low complexity scores, corroborating the interpretation of a minimalist aesthetic based on few recognizable elements.

\section{Conclusions}
In this paper, we examine album covers not only as artistic artifacts but also as vehicles of cultural communication. From this perspective, understanding their evolution over time is crucial, particularly in light of the transformations brought about by digital technologies.

Consistent with previous studies \cite{DiMarco2025,ParadaCabaleiro2024,sigaki2018history}, our analysis, based on multiple measures of visual complexity and object recognition methods, reveals an overall simplification of album covers over time, with notable exceptions in genres such as Metal and Hip Hop. A closer look suggests that more experimental genres tend to increase complexity through abstraction, whereas other genres rely on artists’ images, likely as a promotional tool, resulting in lower visual complexity.

More broadly, investigating the visual complexity of album covers contributes to understanding how cultural products circulate and acquire meaning in the contemporary media environment. Since platforms such as social networks rely heavily on visual cues to capture attention, trends in design and simplification can influence not only aesthetic choices but also the visibility and reception of cultural artifacts. By linking visual complexity to broader processes of cultural communication, this type of study provides valuable insights for media analysis, highlighting how artistic and promotional strategies intersect with the dynamics of digital platforms.

\bibliographystyle{abbrv}
\bibliography{bibliography}

\begin{thebibliography}{10}

\bibitem{adami2002complexity}
C.~Adami.
\newblock What is complexity?
\newblock {\em BioEssays}, 24(12):1085--1094, 2002.

\bibitem{batty2016complexity}
M.~Batty.
\newblock Complexity in city systems: Understanding, evolution, and design.
\newblock In {\em A planner's encounter with complexity}, pages 99--122.
  Routledge, 2016.

\bibitem{Brusila2021}
J.~Brusila, M.~Cloonan, and K.~Ramstedt.
\newblock Music, digitalization, and democracy.
\newblock {\em Popular Music and Society}, 45(1):1–12, Oct. 2021.

\bibitem{DiMarco2025}
N.~Di~Marco, E.~Loru, A.~Galeazzi, M.~Cinelli, and W.~Quattrociocchi.
\newblock Decoding musical evolution through network science.
\newblock {\em arXiv preprint arXiv:2501.07557}, 2025.

\bibitem{donderi2006visual}
D.~C. Donderi.
\newblock Visual complexity: a review.
\newblock {\em Psychological bulletin}, 132(1):73, 2006.

\bibitem{dorochowicz2019relationship}
A.~Dorochowicz and B.~Kostek.
\newblock Relationship between album cover design and music genres.
\newblock In {\em 2019 Signal Processing: Algorithms, Architectures,
  Arrangements, and Applications (SPA)}, pages 93--98. IEEE, 2019.

\bibitem{grebenkina2018edge}
M.~Grebenkina, A.~Brachmann, M.~Bertamini, A.~Kaduhm, and C.~Redies.
\newblock Edge-orientation entropy predicts preference for diverse types of
  man-made images.
\newblock {\em Frontiers in Neuroscience}, 12:678, 2018.

\bibitem{greenfield2024album}
K.~Greenfield and E.~Paintsil.
\newblock Album covers deserve some attention.
\newblock 2024.

\bibitem{yolov8}
G.~Jocher, A.~Chaurasia, J.~Qiu, and et~al.
\newblock {YOLOv8}.
\newblock \url{https://github.com/ultralytics/ultralytics}, 2023.
\newblock Accessed 14th July 2025.

\bibitem{jones1999steve}
S.~Jones and M.~Sorger.
\newblock Covering music: a brief history and analysis of album cover design.
\newblock {\em Journal of popular Music studies}, 11(1):68--102, 1999.

\bibitem{kanstrup2024}
A.~M. Kanstrup et~al.
\newblock Algorithmic ways of seeing: Using object detection to facilitate art
  exploration.
\newblock In {\em CHI '24}, 2024.

\bibitem{lam2002evaluation}
N.~S.-N. Lam, H.-l. Qiu, D.~A. Quattrochi, and C.~W. Emerson.
\newblock An evaluation of fractal methods for characterizing image complexity.
\newblock {\em Cartography and Geographic Information Science}, 29(1):25--35,
  2002.

\bibitem{lang2021}
S.~Lang and B.~Ommer.
\newblock Transforming information into knowledge: How computational methods
  reshape art history.
\newblock {\em Digital Humanities Quarterly}, 15(3), 2021.

\bibitem{li2008introduction}
M.~Li, P.~Vit{\'a}nyi, et~al.
\newblock {\em An introduction to Kolmogorov complexity and its applications},
  volume~3.
\newblock Springer, 2008.

\bibitem{libeks2011you}
J.~Libeks and D.~Turnbull.
\newblock You can judge an artist by an album cover: Using images for music
  annotation.
\newblock {\em IEEE MultiMedia}, 18(4):30--37, 2011.

\bibitem{LopezObj2025}
S.~López and A.~Flexer.
\newblock Zero-shot open set object detection in music album covers.
\newblock 07 2025.

\bibitem{mahon2024minimum}
L.~Mahon and T.~Lukasiewicz.
\newblock Minimum description length clustering to measure meaningful image
  complexity.
\newblock {\em Pattern Recognition}, 145:109889, 2024.

\bibitem{manovich2020}
L.~Manovich.
\newblock {\em Cultural Analytics}.
\newblock MIT Press, 2020.

\bibitem{marin2013examining}
M.~M. Marin and H.~Leder.
\newblock Examining complexity across domains: Relating subjective and
  objective measures of affective environmental scenes, paintings and music.
\newblock {\em PLoS ONE}, 8(8):e72412, 2013.

\bibitem{mcveigh2010itunesart}
C.~McVeigh.
\newblock Creating original cover art for itunes, 2010.

\bibitem{oramas2018}
S.~Oramas, O.~Nieto, M.~Sordo, and X.~Serra.
\newblock Multimodal deep learning for music genre classification.
\newblock {\em Transactions of the International Society for Music Information
  Retrieval}, 1(1):4--21, 2018.

\bibitem{palumbo2015computerized}
L.~Palumbo, R.~Ogden, A.~D. Makin, and M.~Bertamini.
\newblock Computerized measures of visual complexity.
\newblock {\em Behavior Research Methods}, 47(3):895--909, 2015.

\bibitem{ParadaCabaleiro2024}
E.~Parada-Cabaleiro, M.~Mayerl, S.~Brandl, M.~Skowron, M.~Schedl, E.~Lex, and
  E.~Zangerle.
\newblock {Song Lyrics Have Become Simpler and More Repetitive Over the Last
  Five Decades}.
\newblock {\em Scientific Reports}, 14(1):5531, 2024.
\newblock Erratum in: Sci Rep. 2024 May 22;14(1):11712.

\bibitem{PEDRAM2008}
P.~PEDRAM and G.~R. JAFARI.
\newblock Mona lisa: The stochastic view and fractality in color space.
\newblock {\em International Journal of Modern Physics C}, 19(06):855–866,
  June 2008.

\bibitem{redies2007fractal}
C.~Redies, J.~Hasenstein, and J.~Denzler.
\newblock Fractal-like image statistics in visual art: similarity to natural
  scenes.
\newblock {\em Spatial vision}, 21, 2007.

\bibitem{sigaki2018history}
H.~Y.~D. Sigaki, M.~Perc, and H.~V. Ribeiro.
\newblock {History of art paintings through the lens of entropy and
  complexity}.
\newblock {\em Proceedings of the National Academy of Sciences of the United
  States of America}, 115(37):E8585--E8594, 2018.

\bibitem{Taylor1999}
R.~P. Taylor, A.~P. Micolich, and D.~Jonas.
\newblock Fractal analysis of pollock’s drip paintings.
\newblock {\em Nature}, 399(6735):422–422, June 1999.

\bibitem{valensise2021entropy}
C.~M. Valensise, A.~Serra, A.~Galeazzi, G.~Etta, M.~Cinelli, and
  W.~Quattrociocchi.
\newblock Entropy and complexity unveil the landscape of memes evolution.
\newblock {\em Scientific Reports}, 11(1):20022, 2021.

\bibitem{Varnum2021}
M.~E.~W. Varnum, J.~A. Krems, C.~Morris, A.~Wormley, and I.~Grossmann.
\newblock Why are song lyrics becoming simpler? a time series analysis of
  lyrical complexity in six decades of american popular music.
\newblock {\em PLoS ONE}, 16(1):e0244576, 2021.

\bibitem{venkatesan2022does}
T.~Venkatesan, Q.~J. Wang, and C.~Spence.
\newblock Does the typeface on album cover influence expectations and
  perception of music?
\newblock {\em Psychology of Aesthetics, Creativity, and the Arts}, 16(3):487,
  2022.

\bibitem{wang2024complexity}
H.~Wang, C.~Song, and P.~Gao.
\newblock Complexity and entropy of natural patterns.
\newblock {\em PNAS nexus}, 3(10):pgae417, 2024.

\end{thebibliography}



\end{document}